\documentclass{sig-alternate-05-2015}

\usepackage{array}
\usepackage{color}
\usepackage{amsmath}
\usepackage{enumitem}
\usepackage{epstopdf}
\usepackage{url}
\usepackage{bm}
\usepackage{graphicx}
\usepackage{hyperref}
\usepackage{bbm}
\usepackage[tight,footnotesize]{subfigure}
\usepackage{tikz}
\newtheorem{theorem}{Theorem}

\begin{document}
% Copyright
%\setcopyright{acmcopyright}
%\setcopyright{acmlicensed}
%\setcopyright{rightsretained}
%\setcopyright{usgov}
%\setcopyright{usgovmixed}
%\setcopyright{cagov}
%\setcopyright{cagovmixed}

%\CopyrightYear{2017}
%\setcopyright{acmcopyright}
%\conferenceinfo{MM '17,}{October 23-27, 2017, Mountain View, USA}
%\isbn{978-1-4503-3603-1/16/10}\acmPrice{\$15.00}
%\doi{http://dx.doi.org/10.1145/2964284.2964302}

%\acmPrice{\$15.00}

%
% --- Author Metadata here ---
%\conferenceinfo{WOODSTOCK}{'97 El Paso, Texas USA}
%\CopyrightYear{2007} % Allows default copyright year (20XX) to be over-ridden - IF NEED BE.
%\crdata{0-12345-67-8/90/01}  % Allows default copyright data (0-89791-88-6/97/05) to be over-ridden - IF NEED BE.
% --- End of Author Metadata ---

\title{Encoding Bitrate Optimization Using Playback Statistics for HTTP-based Adaptive Video Streaming}

\numberofauthors{4} %  in this sample file, there are a *total*
\author{
%% You can go ahead and credit any number of authors here,
%% e.g. one 'row of three' or two rows (consisting of one row of three
%% and a second row of one, two or three).
%%
%% The command \alignauthor (no curly braces needed) should
%% precede each author name, affiliation/snail-mail address and
%% e-mail address. Additionally, tag each line of
%% affiliation/address with \affaddr, and tag the
%% e-mail address with \email.
%%
% 1st. author
\alignauthor
Chao Chen\\
\affaddr{Google Inc.}\\
       \affaddr{1600 Amphitheatre Parkway}\\
       \affaddr{Mountain View, California}\\
       \email{chaochen@google.com}
% 2nd. author
\alignauthor
Yao-Chung Lin\\
\affaddr{Google Inc.}\\
       \affaddr{1600 Amphitheatre Parkway}\\
       \affaddr{Mountain View, California}\\
       \email{yclin@google.com}
% 3rd. author
\and\alignauthor
Anil Kokaram\\
\affaddr{Department of Electronic and Electrical Engineering, Trinity College Dublin,}\\
       \affaddr{College Green, Dublin, Ireland}\\
       \email{anil.kokaram@tcd.ie}
% 4th. author
\alignauthor
Steve Benting\\
\affaddr{Google Inc.}\\
       \affaddr{1600 Amphitheatre Parkway}\\
       \affaddr{Mountain View, California}\\
       \email{sbenting@google.com}
}
% There's nothing stopping you putting the seventh, eighth, etc.
% author on the opening page (as the 'third row') but we ask,
% for aesthetic reasons that you place these 'additional authors'
% in the \additional authors block, viz.
%\additionalauthors{Additional authors: John Smith (The Th{\o}rv{\"a}ld Group,
%email: {\texttt{jsmith@affiliation.org}}) and Julius P.~Kumquat
%(The Kumquat Consortium, email: {\texttt{jpkumquat@consortium.net}}).}
% Just remember to make sure that the TOTAL number of authors
% is the number that will appear on the first page PLUS the
% number that will appear in the \additionalauthors section.

\maketitle
\begin{abstract}
HTTP video streaming is in wide use to deliver video over the Internet. With HTTP adaptive steaming, a video playback dynamically selects a video stream from a pre-encoded representation based on available bandwidth and viewport (screen) size. The viewer's video quality is therefore influenced by the encoded bitrates. We minimize the average delivered bitrate subject to a quality lower bound on a per-chunk basis by modeling the probability that a player selects a particular encoding. Through simulation and real-world experiments, the proposed method saves 9.6\% of bandwidth while providing the same average delivered video quality comparing with state of the art.
\end{abstract}



%
%  Use this command to print the description
%
\printccsdesc
% We no longer use \terms command


\section{Introduction}
\begin{figure*}
\includegraphics[width=\textwidth]{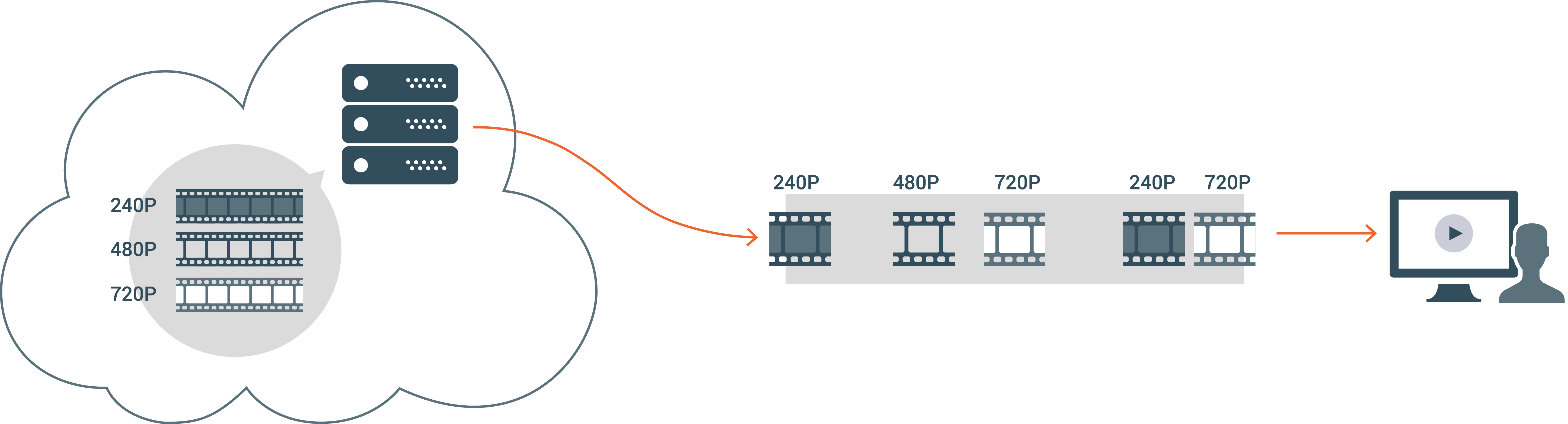}
\caption{Adaptive video streaming over HTTP.\label{fig:dash}}
\end{figure*}

To avoid buffering stalls while streaming videos over communication networks with time-varying throughput, adaptive video streaming protocols based on Hypertext Transfer Protocol (HTTP) have been widely deployed~\cite{seufert2015survey}. Among them, the most popular ones are Dynamic Adaptive Streaming over HTTP (DASH)~\cite{DASH} issued by Moving Picture Experts Group (MPEG) and the HTTP Live Streaming (HLS) protocol~\cite{HLS} proposed by Apple Inc. As illustrated in Fig.~\ref{fig:dash}, with HTTP-based adaptive streaming, videos are encoded into multiple representations at different bitrates and resolutions. Each representation is then partitioned into short video segments. At any moment, the video player dynamically requests the streaming server to send a segment from an appropriate representation: one where the bitrate of the segment is lower than the available network bandwidth capacity and the resolution of the segment fits into the player's viewport.

From the perspective of streaming service providers, e.g., YouTube and Netflix, it is always desirable to deliver maximum video quality with minimum data bitrates. The encoding bitrates and qualities of the representations, to a large extent, determine the delivered video quality and the volume of data traffic. However, it is challenging to find out the optimal encoding bitrates, because the adaptation behavior of the video player affects users' viewing experience and bandwidth cost as well. For example, increasing the encoding bitrate of a representation may not lead to better delivered quality, because that may cause players to switch to other representations with lower bitrates and result in worse video quality presented to viewers. Similarly, decreasing the encoding bitrate of a representation may not reduce the traffic cost, because the players are more likely to select other representations with higher quality, which incurs increased traffic cost.

Rate-quality optimization has been extensively studied in the realm of video compression~\cite{rqo_ref_manage_hevc, benchmark_hevc, avc_rate_control}. These technologies optimize the bit allocation within a single video stream to minimize the encoding bitrate at a given encoding quality. For encoding a single bitstream, the performance of a video codec can be fully characterized by its rate-quality curve. If the rate-quality curve of one codec dominates the rate-quality curve of another codec, the former codec can compress a video at a given quality with a lower bitrate, and thus has better performance. In this paper, we investigate the rate-quality optimization problem for HTTP-based adaptive streaming where players dynamically switch among multiple encoded representations. In this scenario, the average streaming bitrate and delivered quality depend on the encoding bits allocated for all representations.

A method proposed in~\cite{netflix} configures encoding bitrates solely based on the rate-quality curves of representations. The impact of bandwidth and viewport size distributions on the average bitrate and delivered video quality was not considered and is thus sub-optimal for adaptive video streaming.
The optimal selection of DASH representations is also investigated in \cite{kreuzberger2016comparative}, \cite{Li2016Complexity}, and \cite{toni2015optimal}. The problem is formulated as an integer programming that attempts to maximize users' satisfaction given content delivery network (CDN) capacity constraints, content type, and end-user characteristics. Rainer \emph{et al.} proposed an optimization approach for rate adaptation on video streaming when the set of representations are given~\cite{rainer2017statistically}. The idea is to solve a general optimization problem that maximizes delivery quality given download capacity constraint and yield quality upper bound for each streaming session. Then the quality upper bound is folded into another optimization problem that selects which representation to stream. The selected representation minimizes the bandwidth cost while delivering quality close to the upper bound.

In this paper, we aim to minimize the bandwidth cost by optimizing the encoding bitrates for a given set of representations. We consider the large-scale video streaming systems such as {YouTube} or {Netflix} where CDN bandwidth cost dominates the computational cost of encoding. In this scenario, we can spend computational resources to obtain rate-quality curves at all resolutions on a per video chunk basis. It enables us to take into account the unique rate-quality characteristics of each video chunk for encoding bitrate optimization. In addition, we collected real-world playback traces from our video streaming platform and obtained the empirical distributions of estimated client bandwidth and viewport sizes. These empirical distributions enable us to establish a model for the adaptation behavior of players, which is critical to the optimization of encoding bitrates. The optimization step is on a per-chunk basis, i.e. each video chunk can have different representation bitrates instead of a fixed bitrate set per video content type in contrast with the method proposed in~\cite{toni2015optimal}.

We first describe a simple model for the adaptation behavior in players in Section~\ref{sec:method}. In addition to the rate-quality characteristics of the encoded videos at per chunk basis, the model also incorporates the real-world statistics of network bandwidth and the viewport size of players. Using this model, we are able to establish the mapping from the encoding bitrates of encoded representations to the average streaming bitrate and the video quality delivered to viewers. Then, we propose a simple optimization framework to identify the optimal encoding bitrates that minimize the average streaming bitrate, subject to a given lower bound on delivered quality.

We simulated the performance of the proposed method. The results are presented in Section~\ref{sec:sim_result}. The average streaming bitrate can be reduced by 9.45\% to 12.07\%. We then implemented the proposed method in our transcoding system. The experimental results show that the average video traffic is reduced by 9.6\% to 14.37\% without degrading users' Quality-of-Experience. Please see Section~\ref{sec:exp_result} for more detailed results.

\section{Encoding Bitrate Optimization Method}
\label{sec:method}
In this section, we first establish a theorem that characterizes the achievable performance for HTTP-adaptive streaming. Then, we introduce a mathematical model for characterizing the adaptation behavior of video players and formulate the encoding bitrate configurations as an optimization problem. Finally, we describe the overall system implementation.

\subsection{Rate Quality Region}
\label{sec:rate_quality_region}
Let $\mathcal{I}\subset\mathbb{N}$ be the index set of the encoded representations of an input video. Let $\mathcal{V} = \{144, 240, \cdots, 2160\}$ denote the set of supported viewport heights on players. The $i$'th representation is obtained by encoding the input video at bitrate $r_i\in\mathbb{R}$ and resolution $v_i\in\mathcal{V}$. We assume the representations can be ordered by their encoding bitrates and viewport heights in ascending order, i.e., $\forall i < j$ in $\mathcal{I}$, we have $r_i \leq r_j$ and $v_i \leq v_j$. Note that the equality in $v_i \leq v_j$ allows multiple prepresentations per resolution.

Let $q_i$ be the encoding quality of the $i$'th representation. The encoding quality can be measured with any video quality metric such as PSNR or SSIM \cite{SSIM}. Since different representations may have different resolutions, to make their encoding quality comparable, $q_i$ is obtained by first upscaling the representation to the resolution of the input video and then calculating the PSNR or SSIM against the input video.

\begin{figure}
\includegraphics[width=\columnwidth]{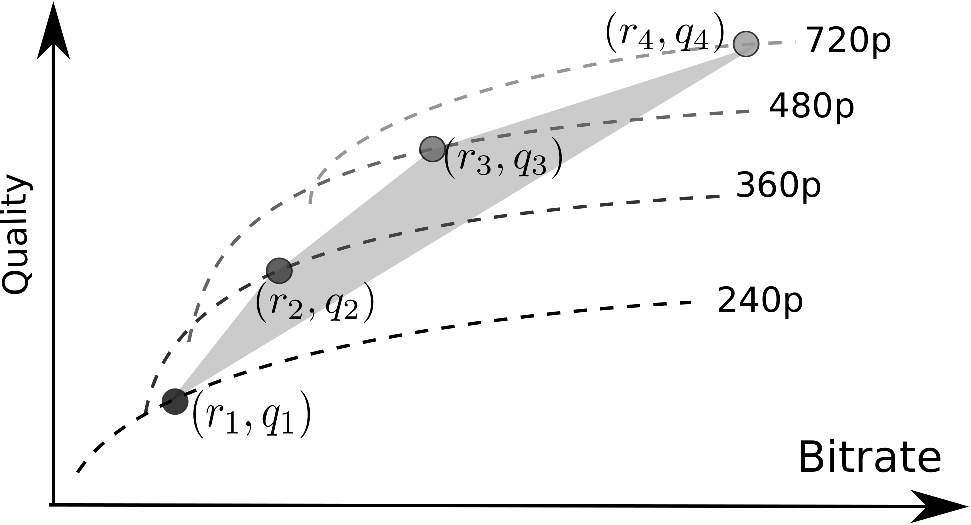}
\caption{Typical rate-quality curves of a video encoded at different resolutions. The achievable rate-quality region is shown by the shaded area.
\label{fig:rq_region}}
\end{figure}

For a given input video and an output resolution $v\in\mathcal{V}$, the encoding quality $q$ at arbitrary encoding bitrate $r$ can be modeled by its rate-quality characteristic function $\mathrm{q}_v(r)$, i.e.,
\begin{align}
q_i = \mathrm{q}_{v_i}(r_i), \forall i\in\mathcal{I}.
\end{align}
The function $\mathrm{q}_v(\cdot)$ depends on the nature of the video content and the compression algorithm adopted by the codec. In Fig.~\ref{fig:rq_region}, we illustrate the typical rate-quality functions of a video corresponding to different encoding resolutions. For each representation, the rate-quality operating point $(r_i, q_i)$ will always fall on the rate-quality curve $\mathrm{q}_{v_i}(\cdot)$. 

Assuming the viewers spent $t_i$ seconds watching representation $i$, the fraction of time viewers spent on representation $i$ is thus
\begin{align}
\label{eq:fraction}
\lambda_i=t_i/\sum_{j\in\mathcal{I}}t_j.
\end{align}
The average streaming bitrate over time is given by
\begin{align}
\label{eq:rate}
\mathrm{R}(\mathbf{r})= \frac{1}{\sum_{j\in\mathcal{I}}t_j}\sum_{i\in\mathcal{I}}t_ir_i=\sum_{i\in\mathcal{I}}\lambda_ir_i = \bm{\lambda}^\intercal\mathbf{r},
\end{align}
where $\mathbf{r}=(r_i:i\in\mathcal{I})$ and $\bm{\lambda} = (\lambda_i:i \in\mathcal{I})$. Similarly, the average quality delivered to users is given by
\begin{align}
\label{eq:quality}
\mathrm{Q}(\mathbf{r}) =\bm{\lambda}^\intercal\mathbf{q},
\end{align}
where $\mathbf{q} = (q_i: i\in\mathcal{I}) = (\mathrm{q}_{v_i}(r_i): i\in\mathcal{I})$ is the vector of encoding qualities.
Because of $\sum_i\lambda_i = 1$ and $\lambda_i \geq 0~\forall i \in\mathcal{I}$, the average bitrate-quality point $\left(\mathrm{R}(\mathbf{r}), \mathrm{Q}(\mathbf{r})\right)$ falls in the convex hull spanned by the encoding rate-quality points $\{(r_i, q_i): i\in\mathcal{I}\}$, which is illustrated in Fig.~\ref{fig:rq_region}. In other words, any rate-quality point outside the convex hull cannot be achieved by an adaptive streaming system. This is summarized in Theorem \ref{theorem:rq_region}.
\begin{theorem}
\label{theorem:rq_region}
For an adaptive video streaming service, the achievable region of average streaming bitrate $R$ and average quality $Q$ is given by the convex hull
\begin{multline}
\label{eq:rq_region}
\mathcal{S}(\mathbf{r}) = \{(R, Q):R = \bm{\lambda}^\intercal\mathbf{r}, Q=\bm{\lambda}^\intercal\mathbf{q},\\ \mathbf{1}^\intercal\bm{\lambda} = 1, \lambda_i \geq 0,~\forall i \in\mathcal{I}\}.
\end{multline}

\end{theorem}

This theorem reveals some of the rationale of the encoding configurations proposed in~\cite{netflix}. This configuration selects the encoding points $\{(r_i, q_i): i\in\mathcal{I}\}$ from the upper boundary of the convex hull spanned by the rate quality curves. It pushes the achievable region $\mathcal{S}(\mathbf{r})$ to the low bitrate and high quality area in the rate-quality space. The corresponding average bitrate tends to be reduced and the average delivered quality tends to be improved. However, the exact position of point $\big(\mathrm{R}(\mathbf{r}),\mathrm{Q}(\mathbf{r})\big)$ depends on $\bm{\lambda}$, i.e., the adaptation behavior of players, which is not considered in ~\cite{netflix}. In the next section, we propose a model to characterize the adaptation behavior of players.

%\begin{figure*}[htp]
%\subfigure[Viewport distribution]{
%\includegraphics[width=\columnwidth]{img/viewport_dist.eps}\label{fig:viewport_dist}
%}
%\subfigure[Bandwidth distribution]{
%\includegraphics[width=\columnwidth]{img/bitrate_dist.eps}
%\label{fig:bitrate_dist_per_viewport}
%}
%\caption{\subref{fig:viewport_dist}: Distribution of the viewport sizes. \subref{fig:bitrate_dist_per_viewport}: The empirical cumulative distribution function under different viewport sizes. The overall distribution is also plotted in the figure as a reference.}
%\end{figure*}

\subsection{Player Model}
\label{sec:player_model}
We model the player-estimated bandwidth and viewport size at players as two stationary random processes $\{\mathsf{R}_t:t\in\mathcal{R}^+\}$ and $\{\mathsf{V}_t:t\in\mathcal{R}^+\}$, respectively. We assume a player selects the streamed representations according to two rules:
\begin{enumerate}
\item The player always requests a representation whose resolution is lower than or equals to player viewport size $\mathsf{V}_t$. This is to save bandwidth by not streaming unnecessary pixels to viewers.
\item Among the representations satisfying the first rule, the player always selects the highest representation whose bitrate is lower than its estimated bandwidth $\mathsf{R}_t$. This is to ensure that the bandwidth is fully utilized while the streamed representation can be smoothly played without stalls.
\end{enumerate}

These two rules are widely followed by video players in practice.
At any moment $t$, a player requests representation $i$ in the following two cases. If the viewport size $\mathsf{V}_t$ equals the resolution of representation $i$, the player requests representation $i$ when the bandwidth $R_t$ is greater than the encoding bitrate $r_i$. If the viewport size $\mathsf{V}_t$ is larger than the resolution of representation $i$, the player requests representation $i$ if $\mathsf{R}_t$ is greater than the bitrate of representation $i$ but is less than that of higher representations. The probability for a player to select representation $i$ is thus
\begin{multline}
\label{eq:viewing_prob_analytical}
\mathrm{p}_i(\mathbf{r}) = \mathbb{P}\left[\mathsf{V}_t = v_i, \mathsf{R}_t > r_i\right]\\
 + \mathbb{P}\left[\mathsf{V}_t > v_i, r_{i+1} \geq \mathsf{R}_t > r_i\right].
\end{multline}
The two terms on the right hand side of \eqref{eq:viewing_prob_analytical} correspond to the probability of the two cases above. We estimated the statistical distributions of bandwidth and viewport sizes from playback statistics and found that the bandwidth distribution does not vary significantly with the viewport size $v$. This is because streaming bandwidth is mainly determined by network conditions when a video is played, which are not related to the viewport size of devices. Therefore, we may assume that viewport size $\mathsf{V}_t$ and bandwidth $\mathsf{R}_t$ are two independent processes. The viewing probability $\mathrm{p}_i(\mathbf{r})$ in \eqref{eq:viewing_prob_analytical} can thus be rewritten as
\begin{multline}
\label{eq:viewing_prob_independence}
\mathrm{p}_i(\mathbf{r})=\mathbb{P}[\mathsf{V}_t=v_i]\mathbb{P}\left[\mathsf{R}_t > r_i\right] \\+ \mathbb{P}\left[\mathsf{V}_t > v_i\right]\mathbb{P}\left[r_{i+1}\geq \mathsf{R}_t > r_i\right].
\end{multline}

Assuming $\mathsf{R}_t$ and $\mathsf{V}_t$ are ergodic random processes, we have
\begin{align}
\label{eq:viewing_prob}
\mathrm{p}_i(\mathbf{r}) =\lim_{\sum_i t_i \rightarrow \infty}\frac{t_i}{\sum_i t_i}=\lambda_i
\end{align}
where $\lambda_i$ is defined in equation \eqref{eq:fraction} as the fraction of time that players stream representation $i$. Substituting \eqref{eq:viewing_prob} into \eqref{eq:rate} and \eqref{eq:quality}, we have
\begin{align}
\mathrm{R}(\mathbf{r}) &=\sum_{i\in\mathcal{I}}r_i\mathrm{p}_i(\mathbf{r}),\label{eq:rate_model}\\
\mathrm{Q}(\mathbf{r}) &=\sum_{i\in\mathcal{I}}q_i\mathrm{p}_i(\mathbf{r})\label{eq:quality_model}.
\end{align}
Substituting \eqref{eq:viewing_prob_independence} into \eqref{eq:rate_model} and \eqref{eq:quality_model}, we obtain the explicit expression for average bitrate and delivered quality in \eqref{eq:explicit_rate} and \eqref{eq:explicit_quality}.

\begin{multline}
\label{eq:explicit_rate}
\mathrm{R}(\mathbf{r}) = \sum_{i\in\mathcal{I}}\Bigg[r_i\bigg(\mathbb{P}\Big[\mathsf{V}_t=v_i\Big]\mathbb{P}\Big[\mathsf{R}_t > r_i\Big]\\ +\mathbb{P}\Big[\mathsf{V}_t > v_i\Big]\mathbb{P}\Big[r_{i+1}\geq \mathsf{R}_t > r_i\Big]\bigg)\Bigg],
\end{multline}
and
\begin{multline}
\label{eq:explicit_quality}
\mathrm{Q}(\mathbf{r}) = \sum_{i\in\mathcal{I}}\Bigg[\mathrm{q}_{v_i}(r_i)\bigg(\mathbb{P}\Big[\mathsf{V}_t=v_i\Big]\mathbb{P}\Big[\mathsf{R}_t > r_i\Big]\\ +\mathbb{P}\Big[\mathsf{V}_t > v_i\Big]\mathbb{P}\Big[r_{i+1}\geq \mathsf{R}_t > r_i\Big]\bigg)\Bigg].
\end{multline}
These expressions can then be used to estimate the corresponding average bitrate and average delivered quality for a given encoding bitrate configuration $\mathbf{r}$.

\subsection{Encoding Bitrate Optimization}
\label{sec:opt_problem}
We propose minimizing the average bitrate, subject to a given lower bound on average delivered quality, by solving the following optimization problem
\begin{equation}
\label{eq:primal_formulation}
\begin{aligned}
& \underset{\mathbf{r}}{\mathrm{minimize}}
& & \mathrm{R}(\mathbf{r})\\
& \text{subject to}
& &\mathrm{Q}(\mathbf{r})\geq\mathrm{Q}_0,
\end{aligned}
\end{equation}
where $\mathrm{R}(\mathbf{r})$ and $\mathrm{Q}(\mathbf{r})$ are given by equation \eqref{eq:explicit_rate} and \eqref{eq:explicit_quality}. The value $\mathrm{Q}_0$ is a desired lower bound for delivered quality.

Here, we target minimizing average bitrate given a constraint on quality. In 2015, video traffic accounted for 70\% of IP network traffic and 55\% of mobile network traffic\cite{VNI2016}. Reducing bandwidth costs is critical for the success of streaming services.

To solve the optimization problem in \eqref{eq:primal_formulation}, gradient-based optimization algorithms such as method of moving asymptotes (MMA)~\cite{MMA} can be employed. It is worth noting that the optimization problem \eqref{eq:primal_formulation} is non-convex because the bitrate cumulative distribution function $\mathbb{P}[\mathsf{R}_t \leq x]$ is not a convex function of $x$ in general. Thus, there is no guarantee that the local optimal solution generated by gradient-based optimization algorithms could achieve a global optimum. However, in practice, we found that the local optimal solutions still provided a significant reduction in average bitrate.

\subsection{System Implementation}
\label{sec:implementation}
The proposed method is integrated in our video processing pipeline that processes and re-encodes ingested videos. An ingested video is first divided into 5-second nonoverlapped chunks. Then we obtain the rate-quality models of each video chunk at all resolutions. For example, we encode a 1080p video chunk into 6 different resolutions including 144p, 240p, 360p, 480p, 720p, and 1080p. For each resolution, we constructed the rate-quality model by sampling rate-quality points from $\mathrm{q}_{v_i}(r)$. Specifically, we encoded a video multiple times using the libx264 codec. Each time we applied a different Constant Rate Factor (CRF) sampled from 5 to 55 with a step size of 5 so as to cover a wide range of encoding qualities. Then we scaled up each encoded version to 1080p with bicubic scaling filter and calculated the corresponding PSNR against the original video. The rate-quality model $\mathrm{q}_{v_i}(r)$ was approximated using the piece-wise linear function connecting the sampled rate-quality points.

On our video streaming platform, the video players recorded the estimated bandwidth in every few seconds during each playback session. The estimated bandwidth, along with the viewport size is then stored to our backend databases. We collected 1,000,000 such real-world playback traces to obtain the empirical distribution of viewport and estimated bandwidth, i.e., $\mathbb{P}[\mathrm{V}_t]$ and $\mathbb{P}[\mathrm{R}_t]$.

The optimizer described in Sec.~\ref{sec:opt_problem} takes into account the rate-quality curves, the empirical distributions of viewport and estimated downlink bandwith, and the default delivered quality $\mathrm{Q}_0$ based on the default settings. It then runs the MMA solver in the NLopt library~\cite{nlopt} to obtain the optimized bitrates. Finally, the video chunks will be encoded into the representations using the optimized bitrates, and muxed into DASH or HLS formats for streaming.

In the next section, we evaluate the gain of the optimized encoding bitrate configurations via numerical simulations.

\section{Numerical Simulations}
\label{sec:sim_result}
A set of 1000 1080p videos, the content of which covers a wide-range of spatio-temporal complexities, is randomly selected to run the simulation. The lengths of the selected videos ranged from 1 minute to 20 minutes. Every videos are processed as described in Sec.~\ref{sec:implementation}.

We first evaluated the performance of the proposed method against a baseline encoding parameter configuration where a fixed CRF of 23 is applied to all representations. Here, we chose a CRF of 23 because it is the default CRF in ffmpeg. We denote by $r_i^0$ the baseline encoding bitrate for representation $i$ and let $\mathbf{r}_0=\left(r_i^0: i\in\mathcal{I}\right)$. We set $\mathrm{Q}_0 = \mathrm{Q}(\mathbf{r}_0)$ as the lower bound for delivered quality. Then we employed the MMA solver in the NLopt library \cite{nlopt} to find the optimal encoding bitrates $\mathbf{r}^*$ and the corresponding average streaming bitrate $\mathrm{R}(\mathbf{r}^*)$ in \eqref{eq:primal_formulation}.

\begin{figure}
\includegraphics[width=\columnwidth]{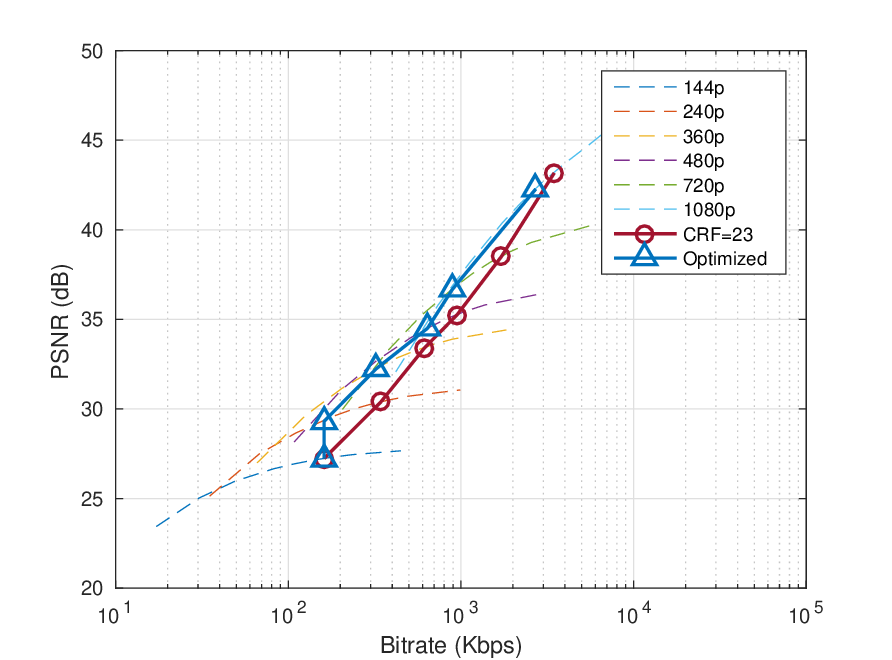}
\caption{The encoding bitrates when CRF is fixed to 23 ($\bigcirc$) and the corresponding encoding bitrates optimized by the proposed method ($\bigtriangleup$) for an example video.\label{fig:opt_vs_fixcrf}}
\end{figure}

\begin{figure}
\includegraphics[width=\columnwidth]{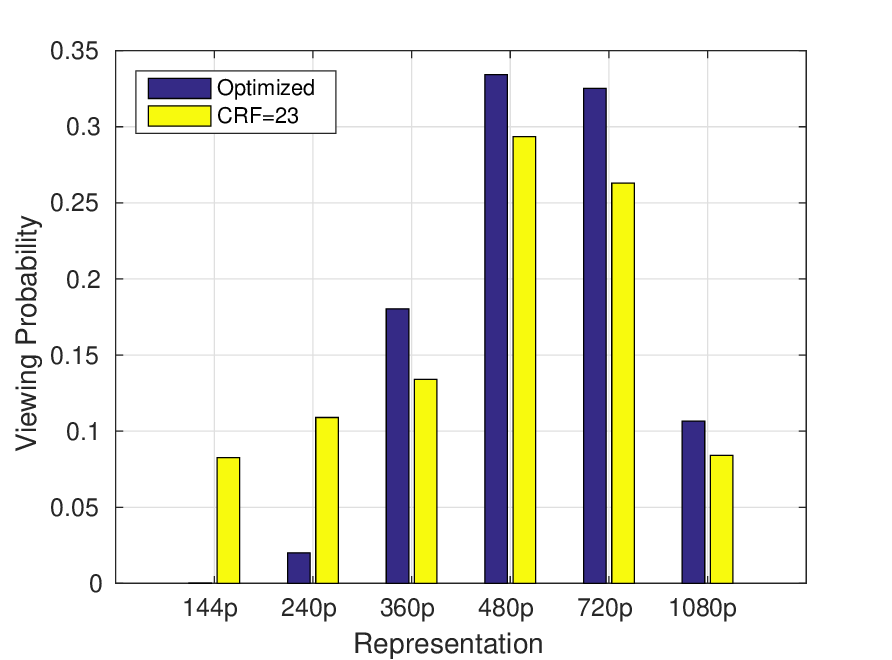}
\caption{The viewing probability distribution of an example video estimated by the proposed player model.\label{fig:opt_vs_fixcrf_view_prob}}
\end{figure}

On the 1000 test videos, we found that the proposed algorithm can reduce the average streaming bitrate by 12.07\%. We plot the encoding configurations of an example video in Fig.~\ref{fig:opt_vs_fixcrf}. It can be seen that the optimized encoding bitrates are smaller than the encoding bitrates with fixed CRFs. This leads to reduction in encoding bitrate and degradation in encoding quality for all representations. Interestingly, because the delivered quality depends on the viewing probability distribution, the delivered quality of the optimized encoding configuration is kept the same as that of the baseline configuration. In Fig.~\ref{fig:opt_vs_fixcrf_view_prob}, we plot the viewing probability of each representation as predicted by our player model in \eqref{eq:viewing_prob_analytical}. It is seen that the optimized encoding configuration tend to cause the player to spend more time streaming higher representations, thereby compensating for the loss in delivered quality due to reduced encoding bitrates.

\begin{figure}
\includegraphics[width=\columnwidth]{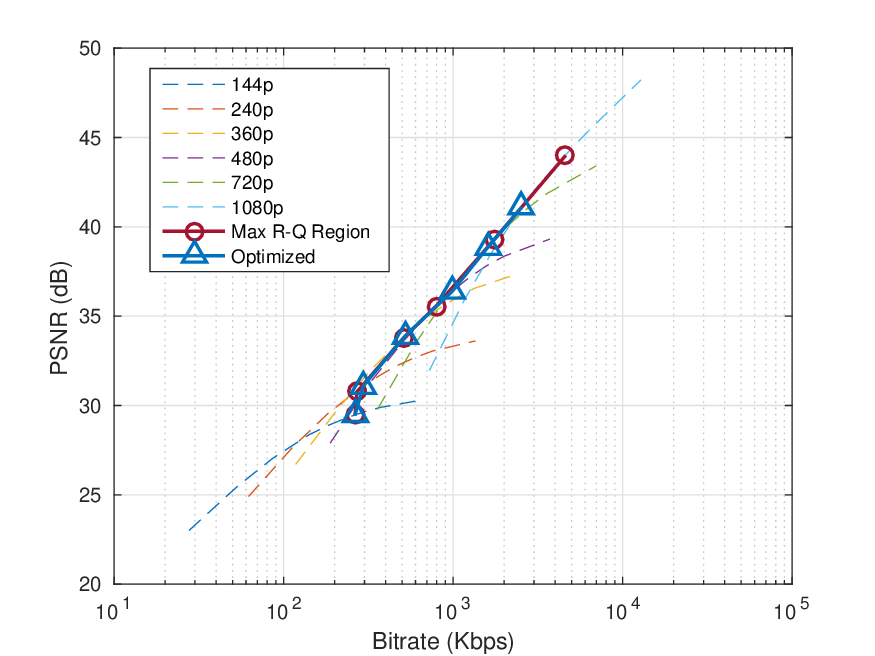}
\caption{The encoding parameter configurations that maximized the achievable rate quality region ($\bigcirc$) and the corresponding encoding bitrates optimized by the proposed method ($\bigtriangleup$) for an example video.\label{fig:opt_vs_netflix}}
\end{figure}

\begin{figure}
\includegraphics[width=\columnwidth]{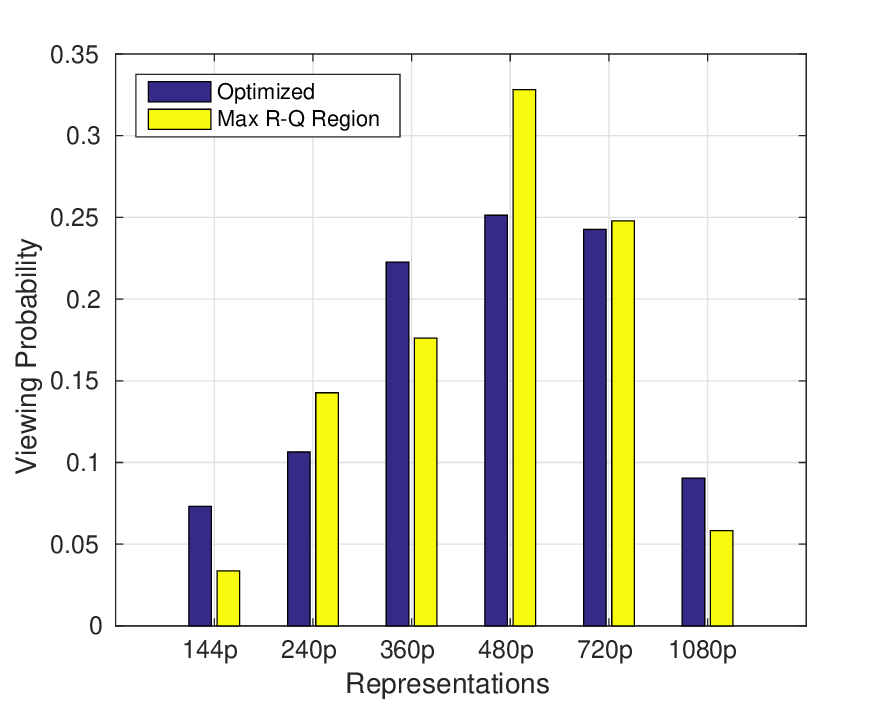}
\caption{The viewing probability distribution of an example video estimated by the proposed player model.\label{fig:opt_vs_netflix_view_prob}}
\end{figure}

We also compared the proposed solution with another baseline configuration. In this baseline configuration, we fixed the encoding CRFs of the 144p and 1080p representations to be 23. Then we selected the encoding bitrates for the other representations such that the achievable rate-quality region given in \eqref{eq:rq_region} was maximized. This is similar to the method proposed in \cite{netflix}. On the 1000 test videos, we found the proposed solution could reduce the average streaming bitrate by 9.45\%. Fig.~\ref{fig:opt_vs_netflix} illustrated the baseline and the optimized encoding configurations of an example video. It can be seen that the optimized encoding bitrate is higher than the baseline configurations in low resolution presentations, including 240p, 360p, and 480p. For 720p and 1080p representation, the optimized encoding bitrates are lower than that of the baseline. As shown in Fig.~\ref{fig:opt_vs_netflix_view_prob}, with the optimized configuration, players would spent more time on the 1080p representations, preserving delivered quality.

In the next section, we validate the effectiveness of our encoding bitrate configration via experiments on our video streaming platform.

\section{Experimental Results}
\label{sec:exp_result}
We randomly selected 5,000 1080p videos for a real-world experiment. The selected videos were 1 minute to 20 minutes long. Two pairs of treatments were applied, each containing a baseline encoding bitrate configuration and the corresponding optimized configuration. As in our simulations, the first pair of treatments used the default ffmpeg CRF of 23 as the baseline. In the second pair, the baseline configuration fixed the encoding CRF for 144p and 1080p at 23. The bitrates of other representations were configured such that the achievable rate quality region was maximized.

For each treatment, the bitrate configurations were applied on every 5-second chunks of the videos in order to incorporate the variations in the spatial-temporal characteristics of videos. The playback statistics, which included total watch time and average video streaming bitrates, were collected to evaluate performance. In the following, we will first report the bitrate changes in each treatment pair, and then summarize the overall playback statistics.

\begin{figure}
\subfigure[]{
\includegraphics[width=\columnwidth]{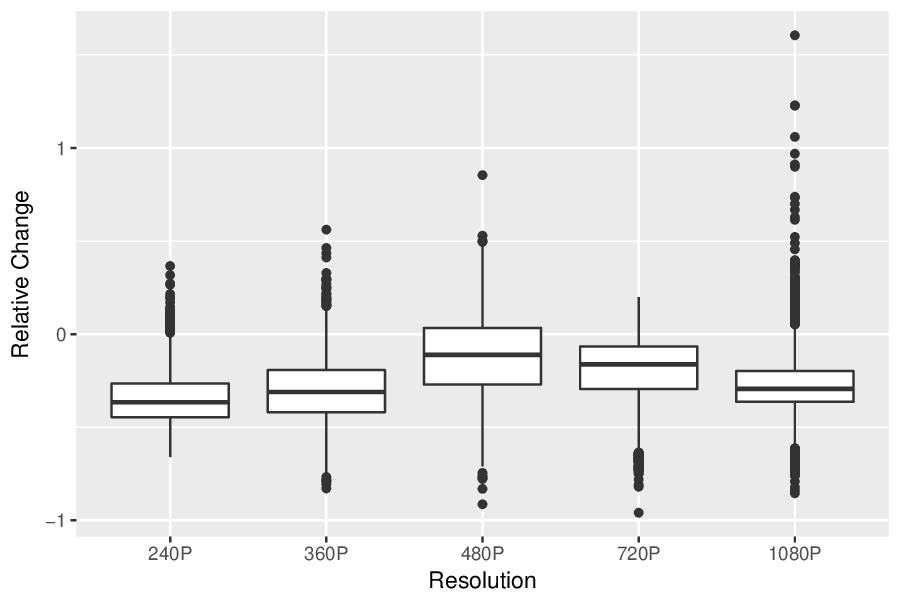}
\label{fig:bitrate_diff_ratio_fixedcrf}
}
\subfigure[]{
\includegraphics[width=\columnwidth]{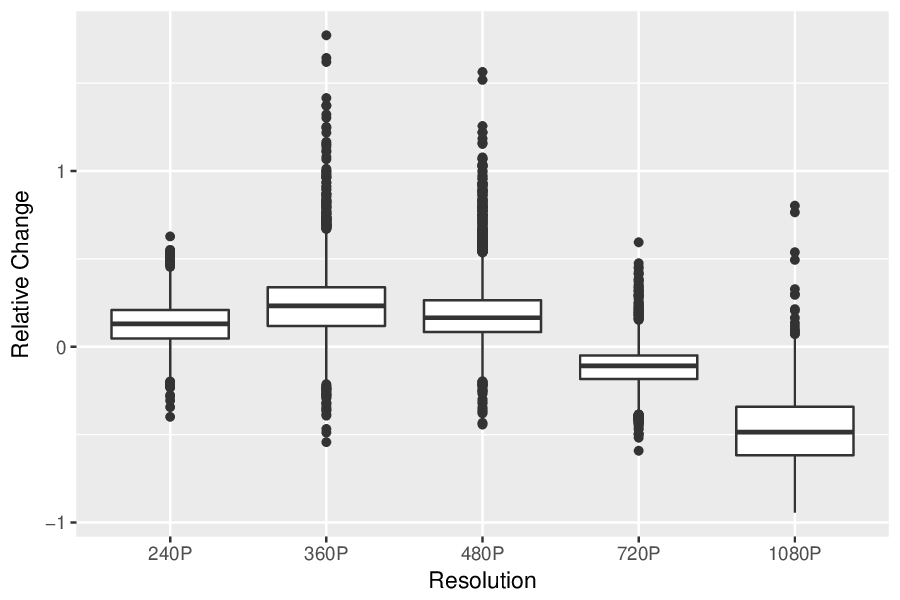}
\label{fig:bitrate_diff_ratio_maxregion}
}
\caption{Relative encoding bitrate changes per resolution under the two experimental settings: \subref{fig:bitrate_diff_ratio_fixedcrf} Fixed CRF = 23 as the baseline. \subref{fig:bitrate_diff_ratio_maxregion} Maximizing rate-quality region as the baseline.}
\label{fig:bitrate_diff_ratio}
\end{figure}

% Bitrate change
We define the relative change in encoding bitrate as $\frac{\mathrm{r}^*}{\mathrm{r}_0} - 1$, where $\mathrm{r}^*$ and $\mathrm{r}_0$ are the encoded bitrates of the optimized configuration and the corresponding baseline configuration, respectively.
Fig.~\ref{fig:bitrate_diff_ratio_fixedcrf} shows the boxplot of relative encoded bitrate changes at different resolutions for the first pair of treatments. It can be seen that the proposed method selects lower encoding bitrates for almost all resolutions. This is especially true for 240P, where the median of the relative change was -36.5\%. Fig.~\ref{fig:bitrate_diff_ratio_maxregion} shows the relative encoding bitrate changes against the baseline that maximizes the achievable rate-quality region. In this case, the optimizer increases the median bitrates of 240p, 360p, and 480p by 13.0\%, 23.2\% and 16.5\%,  respectively. The median bitrates of 720p and 1080p are reduced by 10.8\%, and 48.5\%, respectively.

\begin{figure}
\subfigure[]{
\includegraphics[width=\columnwidth]{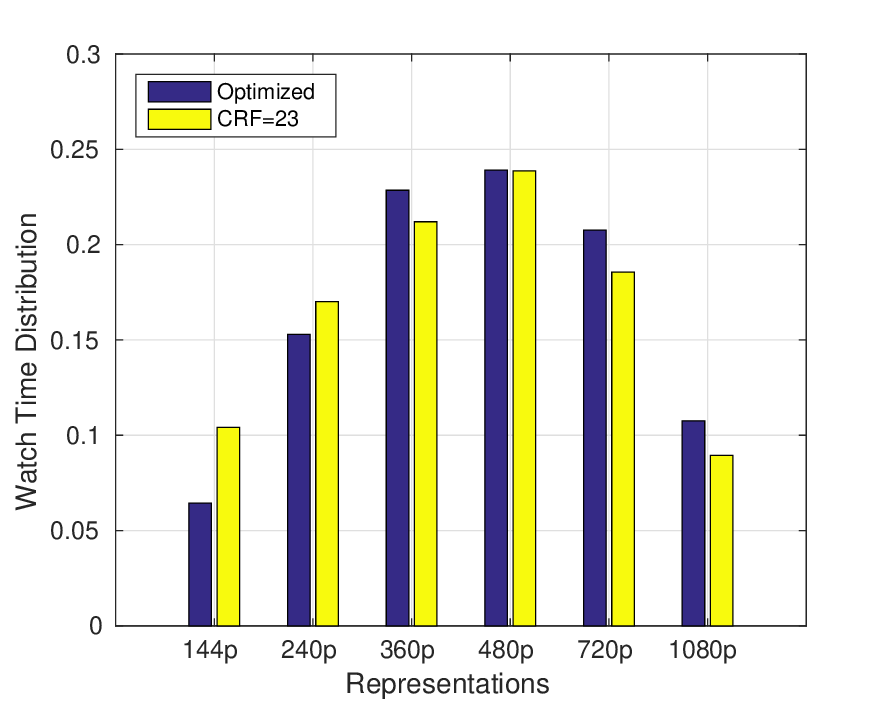}
\label{fig:playtime_dist_fixedcrf}
}
\subfigure[]{
\includegraphics[width=\columnwidth]{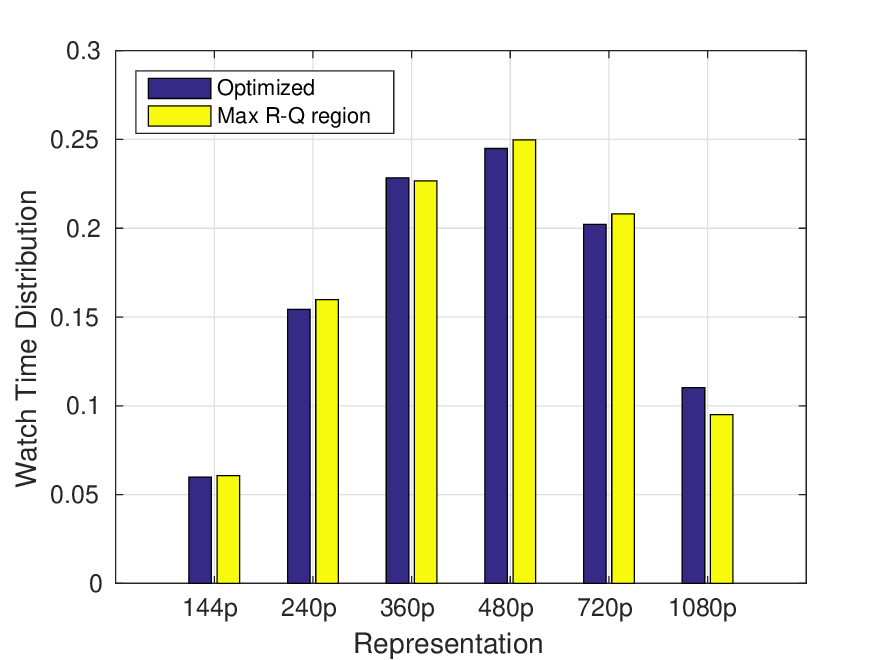}
\label{fig:playtime_dist_maxregion}
}
\caption{Watch time distribution at different resolutions under the two experiment settings: \subref{fig:playtime_dist_fixedcrf} Fixed CRF as the baseline. \subref{fig:playtime_dist_maxregion} Maximizing rate-quality region as the baseline}
\label{fig:playtime_dist}
\end{figure}

% Watch time distribution
From the collected statistics, the changes in bitrate affected the distribution of watch time across resolutions. Fig.~\ref{fig:playtime_dist} compares watch time distributions at different resolutions. Fig.~\ref{fig:playtime_dist_fixedcrf} shows that, for the first pair of treatments, the watch time of the optimized configuration generally shifts towards higher resolutions. For the second pair of treatments, watch time shifts are observed as well. As can be seen in Fig.~\ref{fig:playtime_dist_maxregion}, the watch time of the 240p representation is slightly reduced while that of the 1080p representations is increased.

% Summary.
We calculated the average streaming bitrate and observed that the proposed method saves 14.37\% in average bitrate against the configuration using a fixed CRF of 23, and 9.65\% against the configurations that maximize the rate-quality region.

The optimized encoding bitrate configurations is designed to reduce average streaming bitrate without affecting the quality of experience of viewers. Table.~\ref{tab:expected_rate_quality_1} and Table.~\ref{tab:expected_rate_quality_2} summarize the total watch time and average delivered PSNR of each configuration. Measured quality loses 0.20 dB in PSNR against the configuration using a fixed CRF of 23, and gains 0.05 dB against the configurations that maximize the rate-quality region. For both treatment pairs, we conducted two-sided log-transformed t-tests at 95\% confidence on watch time and average delivered PSNR with. There is no significant changes in both metrics.

We also measured the representation switching rate, the join latency that is the latency from the moment when a playback request is sent to the moment when the video starts to play and the mean time between rebufferings. As shown in Table.~\ref{tab:expected_rate_quality_1}, comparing with the configuration using a fixed CRF of 23, our method reduces the rate switching rate and initial delay significantly. This can improve the overall QoE of users. The proposed method also increased the mean time between rebuffering events but the improvement is statistically insignificant. Comparing with the configurations that maximize the rate-quality region, our method reduced the rate switching rate slightly. Its impact on initial delay and mean time between rebufferings is statistically insignificant.

\begin{table*}[ht]
\centering
\begin{tabular}{lrrrr}
  \hline
  Metrics & CRF=23 & Optimized & Gain & Statistical Significance \\
  \hline
  Watch Time & 178,201 & 177,499 & -0.39\% & insignificant\\
  Average Quality & 37.93 & 37.73 & -0.53\% & insignificant \\
  Normalized Average Bitrate & 1 & 0.855 & -14.37\% & significant \\
  Normalized Adaptive Switch Rate &1 & 0.8699 & -13.01\% & significant \\
  Normalized Join Latency & 1 & 0.9791 & -2.09\% & significant \\
  Normalized Mean Time Between Rebuffers & 1 & 0.9904 & 0.96\% &insignificant \\
  %Relative Average Bitrate & 1421 & 1216\\
   \hline
\end{tabular}
\caption{Performance comparing with the baseline where CRF of 23 is applied to all representations. Watch Time: total watch time in hours, Average quality: measured quality in scaled PSNR (dB). Normalized Average Bitrate: measured average bitrate divided by the average bitrate of the baseline. Adaptive switch rate: the number of representation switches per hour of video playbacks. Join latency: the latency from the moment when a playback request is sent to the moment when the video starts to play. Mean time between rebuffering: the average time between two playback rebuffering events.}
\label{tab:expected_rate_quality_1}
\end{table*}

\begin{table*}[ht]
\centering
\begin{tabular}{lrrrr}
  \hline
  Metrics & Baseline & Optimized & Gain & Statistical Significance \\
  \hline
  Watch Time &177,265 & 177,797 &  0.30\% & insignificant\\
  Average Quality & 37.49 & 37.54 &  0.13\% & insignificant\\
  Normalized Average Bitrate & 1 & 0.903 & -9.65\% & significant\\
  Normalized Adaptive Switch Rate &1 & 0.9924 & -0.76\% & significant \\
  Normalized Join Latency & 1 & 1.0007 & 0.07\% & insignificant \\
  Normalized Mean Time Between Rebuffers & 1 & 0.9926 & -0.74\%& insignificant\\
  %Average Bitrate & 1347 & 1217\\
   \hline
\end{tabular}
\caption{Performance comparing with the baseline where bitrates are configured to maximize the rate quality region. Watch Time: total watch time in hours. Average quality: measured quality in scaled PSNR (dB). Normalized Average Bitrate: measured average bitrate divided by the average bitrate of the baseline.}
\label{tab:expected_rate_quality_2}
\end{table*}

\section{Conclusions and Future Work}
\label{sec:conclusion}
We propose a mathematical model for the adaptation behavior of players in HTTP-based video streaming. In addition to the rate-quality characteristics of videos, the model also incorporates the statistical distribution of available\\ bandwidth and viewport sizes of players. Based on the model, we implemented a method to optimize the encoding bitrates of the video representations. Both numerical simulations and experimental results demonstrated that the proposed method can save 9.6\% on the average video streaming bandwidth without degrading users' quality of experience or average video delivered quality.

The optimization method presented in this paper is based on global bandwidth and viewport size distributions. However, a video might be popular in a certain geographic region where the bandwidth/viewport distributions differ from the global ones. As part of future work, we will investigate potential gains of incorporating local bandwidth and viewport distributions to our approach.
\bibliographystyle{abbrv}
\bibliography{sigproc}  % sigproc.bib is the name of the Bibliography in this case
% ACM needs 'a single self-contained file'!
%
\end{document}